\documentclass[preprint,1p,times]{elsarticle}

\usepackage{amssymb}
\usepackage{amsthm}
\usepackage{amsmath}
\usepackage{mathrsfs}
\usepackage{xcolor}
\usepackage{hyperref}
\usepackage{babel}
\usepackage{bm}
\usepackage{graphicx}
\usepackage{url}
\usepackage{booktabs}

\usepackage{algorithm}  
\usepackage{algpseudocode}  
\usepackage{amsmath}

\journal{MDPI}
\begin{document}

\begin{frontmatter}
\title{Using R for teaching and research}

\author{M. Isabel Parra}\ead{mipa@unex.es}
\author{Eva L.  Sanjuán}\ead{etlopez@unex.es}
\author{M.  Carmen Robustillo}\ead{mcrobustillo@unex.es}
\author{Mario M. Pizarro\corref{cor}}\ead{mariomp@unex.es}\cortext[cor]{Corresponding author}

\address{Department of Mathematics, University of Extremadura}
\date{Submitted to Axioms}

\begin{abstract}
\texttt{R} is a language and environment for statistical computing and graphics, which provides a wide variety of statistical tools (modeling, statistical testing, time series analysis, classification problems, machine learning, ...), together with amazing graphical techniques and the great advantage that it is highly extensible. Nowadays, there is no doubt that it is the software par excellence in statistical courses for any level, for theoretical and applied subjects alike. Besides, it has become an almost essential tool for every research work that involves any kind of analysis or data visualization. Furthermore, it is one of the most employed programming languages for general purposes. The goal of this work is helping to share ideas and resources to improve teaching and/or research using the statistical software \texttt{R}. We will cover its benefits, show how to get started and where to locate specific resources, and will make interesting recommendations for using \texttt{R}, according to our experience. For the classroom we will develop a curricular and assessment infrastructure to support both dissemination and evaluation, while for  research we will offer a broader approach to quantitative studies that provides an excellent support for work in science and technology. 
\end{abstract}


\end{frontmatter}
\section{Introduction}
There is wide evidence of the necessity of statistical knowledge in the field of actual sciences. For this reason, Statistics occupies a permanent place in the curricula degrees of a very diverse nature, either with the purpose of allowing us to analyze the data form research or with the aim of enable us to understand the data from published researches. 
A key topic in any science is how to make calculations, because it is essential to use adequate tools. Especially for teaching, there is some controversy about making hand calculations versus using appropriate software. According to our experience as teachers and researchers, making simple hand computations with a small dataset can be useful for truly understanding how the method works, but some examples should be enough. However, employing adequate software is mandatory for developing applied and practical skills about statistical methods, and it allows us to start asking and answering really interesting questions. In fact, until a statistical technique is not assimilated in a popular software, it is rarely employed, regardless of its importance or advantages. For these reasons, in the last decades, many families of statistical packages have been developed (as Matlab, Minitab, PSPP, R, SAS/STAT, SPSS, Stata, Statgraphics, Statistica, SLSTAT), which arises the question of deciding which one to employ. The purpose of this work is not to make a comparative analysis between them, but to show the \texttt{R} framework, with recommendations about how to use it. Our extensive experience has driven us to choose it, because it has shown to be of great utility, both for teaching and research, and it has many advantages that we will describe below. 

\texttt{R} was created in 1992, but announced to the public in 1993. In 1995, Martin Mächler, of the Federal Institute of Technology in Zurich persuades Ross and Robert to use the GNU license to make \texttt{R} free software, and it becomes part of the project from 1997. However, until February 29, 2000, the software was not considered complete and stable enough to release version 1.0. \texttt{R} \citep{ref-url1} is a GNU project, similar in appearance to the \texttt{S} language but the underlying implementation and semantics are derived from Scheme \cite{inicio}, developed at Bell Laboratories, available as free software under the terms of the Free Software Foundations's GNU General Public License in source code form. It compiles and runs on a wide variety of UNIX platforms and similar systems as FreeBSD or Linux, Windows and MacOS or even in platforms as improbable as a PlayStation.

The growth in the number of \texttt{R} users in recent years indicates that researchers around the world are either using \texttt{R} or will use it at some point. Despite of its initial design, focused on a very precise objective (data analysis), it appears within the top ranks of programming languages for general purposes. For example, it appeared in 13th position in TIOBE Index for May 2023; 7th position in Popularity of Programming Language, 8th position in IEE Top Programming Languages 2022 or 12th in RedMonk.

There are several advantages of using \texttt{R} for statistical computing, apart from its open and free source. The benefits of \texttt{R} for an introductory student and instructors are many and varied. \texttt{R} is a high-level programming language and environment for statistical computing and graphics which allows users to perform a wide array of sophisticated operations with only a few, and often intuitive, lines of code, especially suitable for the classroom. Even for students and users with little programming background, \texttt{R} can be easy and intuitive to learn. In addition, statistical modelling and analysis can be performed by executing instructions that use pre-programmed functions, which make more accessible   for an applied researcher to use rather than coding its own functions. \texttt{R} has a rich variety of user-contributed packages, including many teaching-related packages, for generating advanced and customizable graphs (R Graph Gallery  \citep{ref-url2}), ... and an excellent built-in help system and various online help pages including mailing list and boards.

Many statisticians and data scientists use \texttt{R} with the command line. Rather than pointing and clicking in a graphical environment, they write code statements to ask \texttt{R} to do something. This has the advantage of providing a record of what was done, which is increasingly required as part of a publication submission, and allows for peer review of the work undertaken. However, the command line can be quite daunting to a beginner of \texttt{R}. Fortunately, there are many different graphical user interfaces (GUI), many of which offer Integrates Development Environment (IDE) facilities, that help to flatten the learning curve.

%We’ve restricted this group test to software that’s released under an open source license, and offers Integrated Development Environment (IDEs) facilities.
\section{Brief introduction to \texttt{R}}
\begin{figure}[H]
%\begin{adjustwidth}{-\extralength}{0cm}
\centering
\includegraphics[width=15cm]{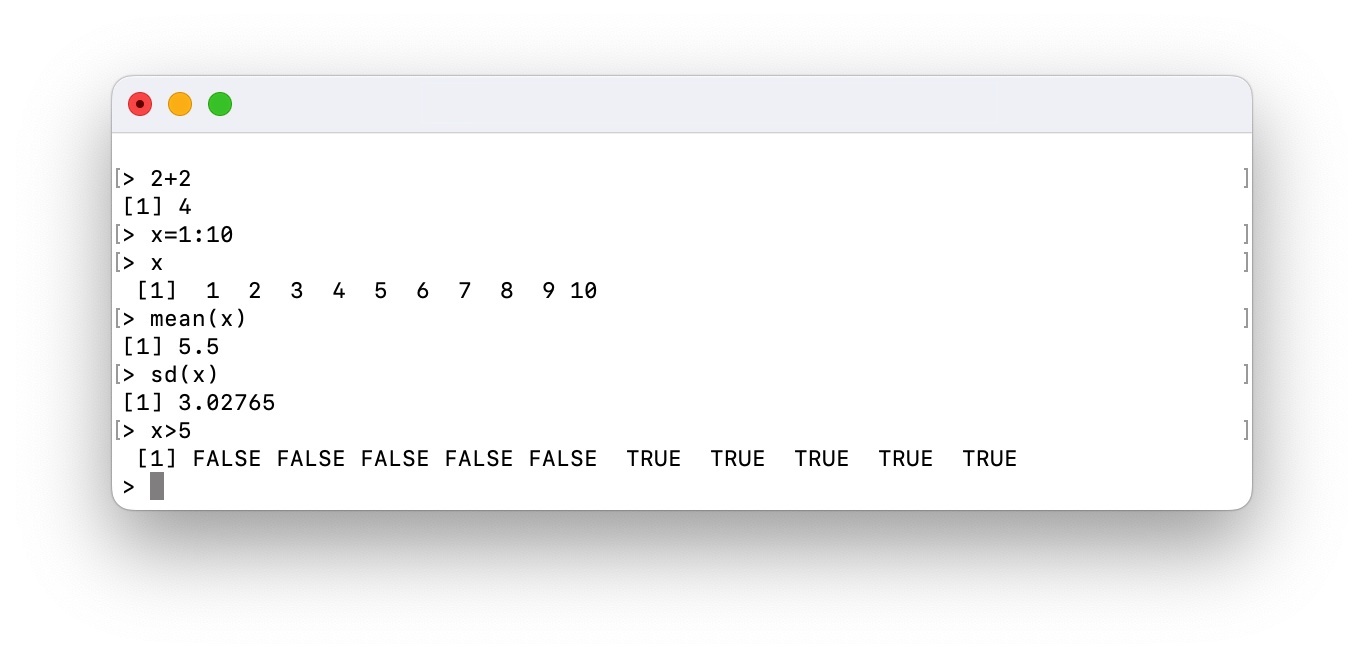}
%\end{adjustwidth}
\caption{R used interactively from terminal.\label{figc}}
\end{figure}

After \texttt{R} is \href{https://cran.r-project.org/}{\textcolor{cyan}{\underline{downloaded}}} and installed, we simply need to find and launch it from our Applications folder. As \texttt{R}  is a command line driven program, the users enter commands at the prompt (\texttt{>}), which are executed one at a time (Figure \ref{figc}). \texttt{R} works with objects, which store data, and commands that can operate on them. A single number or string of text is the simplest object, which is simply a vector of length one, because there is no distinct object type for a scalar. 

Values are assigned to name variables with an arrow (\texttt{<-} or \texttt{->}) or an equal sign (\texttt{=}). \texttt{R} makes calculations using arithmetic operators, comparisons and logical operations.

The workspace is our current \texttt{R} working place and it includes the user-defined objects (vectors, matrices, data frames, list and functions). When we finish an \texttt{R} session, we can save an image of the current workspace that is automatically reloaded the next time \texttt{R} is started.

\subsection{Entering or reading data}
We can either type the data manually or read it from a file, which is fairly simple, because it offers options to import many file types, from CSVs to databases with commands as \texttt{read.table}, \texttt{scan}, \texttt{data} or \texttt{attach}, depending on our situation. 

\texttt{R} allows the following {\it{data structures}}:
\begin{itemize}

\item The basic structure in \texttt{R} is the {\bf{vector}} with all elements of the same type (numeric, character or logical). Most arithmetic operations are vectorized, i.e. the operations are applied element-wise. When one operand is shorter than the other, the first one is recycled, i.e. the values from the shortest one are reused in order to have the same length as the largest, and a warning is printed. 

\item {\bf{Matrices}} are two dimensional vectors created with the \texttt{matrix} function.

\item {\bf{Data frames}} are the most flexible data structure used to store datasets in spreadsheet-like tables. Usually, the observations are the rows and the variables are the columns.

\item {\bf{Lists}} are very powerful data structures, consisting of ordered sets of elements, that can arbitrary store different types of \texttt{R} objects.

\end{itemize}

A distinctive feature of \texttt{R} is that some functions can make data structures to be coerced (i.e. internally converted) from one type to another automatically when needed, or the user can employ specific functions for it (like \texttt{as.vector}, \texttt{as.matrix}...).
%\end{document}

\subsection{Loops and conditionals}

\texttt{R} allows the implementation of control-flow constructs. The most commons are:

\noindent \texttt{for(var in seq) expression}

\noindent \texttt{while(condition) expression}

\noindent \texttt{if(condition) expression} 

\noindent \texttt{if(condition) cons.expression  else  alt.expression}

\noindent They function in much the same way as control statements in any Algol-like language. 

\begin{figure}[H]
%\begin{adjustwidth}{-\extralength}{0cm}
\centering
\includegraphics[width=10cm]{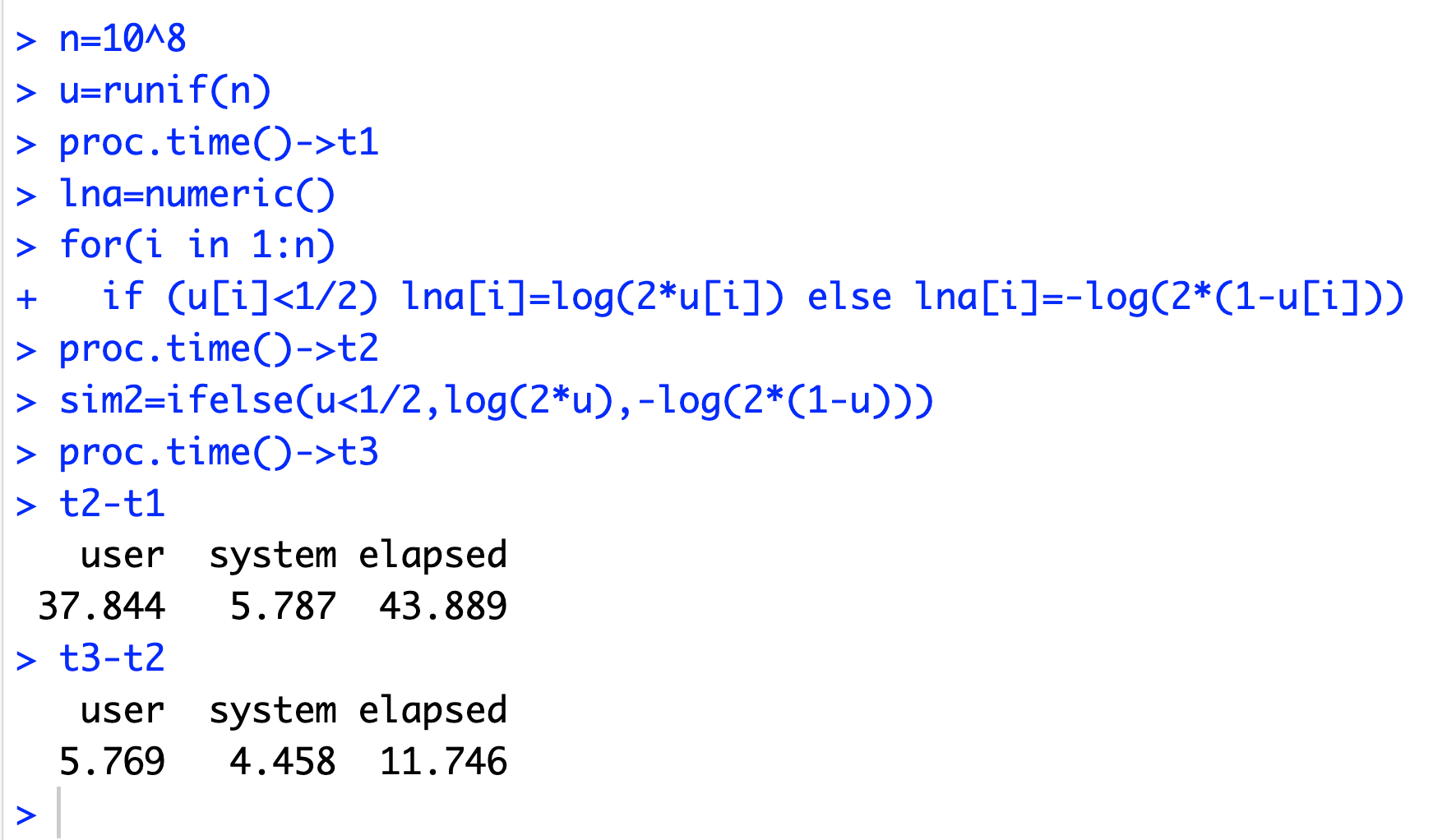}
%\end{adjustwidth}
\caption{The codes compares the simulation of a sequence of $10^8$ values from a Laplace distribution with parameters 0 and 1, by combining \texttt{for} and \texttt{if} and by using \texttt{ifelse}, by means of calculating the computation times for both procedures. \label{ifelse}}
\end{figure}

The \texttt{ifelse} function presents an advantage over the standard \texttt{if} statement, because it combines element-wise operations (vectorized) and filtering with a condition that is evaluated, which involves greater efficiency. An example of use can be seen in the Figure \ref{ifelse}. We can appreciate the great reduction in computation times. 

\subsection{Geting help}
There is a comprehensive built-in help system which can be used with any of the following command:
\begin{table}[h]
\caption{Help and documentation.\label{help}}
	%\begin{adjustwidth}{-\extralength}{0cm}
%\newcolumntype{C}{>{\centering\arraybackslash}X}
\begin{tabular}{cc}
\toprule 
\texttt{help.start()} & general help html based \\
\midrule
\texttt{help(nombre\_function)} & description of the function, possible arguments\\
\texttt{?nombre\_function} &  and values used as default, references, and usage examples\\
\midrule
\texttt{??string}  & show all the topics matching ``string'' \\
\midrule
\texttt{apropos("nombre")}& list all functions containing string ``nombre''\\
\midrule
\texttt{example(nombre\_function)} & show some examples of how the function can be used\\
\bottomrule
\end{tabular}
	%\end{adjustwidth}
\end{table}

\subsection{Programing}
\texttt{R} is an interpreted language (like Java or Python), and not a compiled language (like Fortran, Pascal, C, C++). Commands typed on the keyboard are executed directly without the need to build an executable file, making easy to analyze complex data. It supports procedural programming with functions and, for some functions, object-oriented programming with generic functions. Due to its \texttt{S} heritage, \texttt{R} uses its syntax to represent data and code and has stronger object-oriented programming features than most statistical computing language. Advanced users can write C, C++. Java, Python or NET code to manipulate \texttt{R} objects directly.
\subsubsection{\texttt{R}-functions}
Almost everything in \texttt{R} is done through functions. A function is a set of statements to perform a specific task. \texttt{R} has a large number of built-in functions, addressing a wide variety of problems, which can be directly called. Most of them are coded in \texttt{R}, although to increase efficiency, there are functions coded in low level languages like C or Fortran.

The user can easily create their own functions. It may accept arguments or parameters (one, several or an indefinite number) and it can return one or more values (or not). The basic syntax is as follows: 
\begin{center}
\texttt{function\_name <- function (arg1, arg2, ...) \{ function body \}}
\end{center}
The parts of a function are:
\begin{itemize}
\item {\it{Function name}}, which is stored in \texttt{R}.
\item {\it{Optional arguments}}, which can have default values, are placeholders. When a function is called, it passes a value to the argument without almost default values.
\item {\it{Function body}} contains a collection of declarations that define what the function does.
\item {\it{Return value}} is the last evaluated expression.
\end{itemize}
Functions are first-class objects and can be manipulated in the same way as data objects, making it easy to do metaprogramming that allows multiple dispatch. Function arguments are passed by values and are only evaluated when used, not when the function is called. A generic function can act differently depending on the classes of the arguments passed to it, because it passes the specific method implementation for that particular class. 

Note that by using built-in functions, the only thing you need to worry about is how to effectively communicate the correct input arguments and manage the return values.

{\noindent \bf{¿How to obtain the code of a function?}}

\begin{figure}[htp]
%\begin{adjustwidth}{-\extralength}{0cm}
\centering
\includegraphics[width=\textwidth]{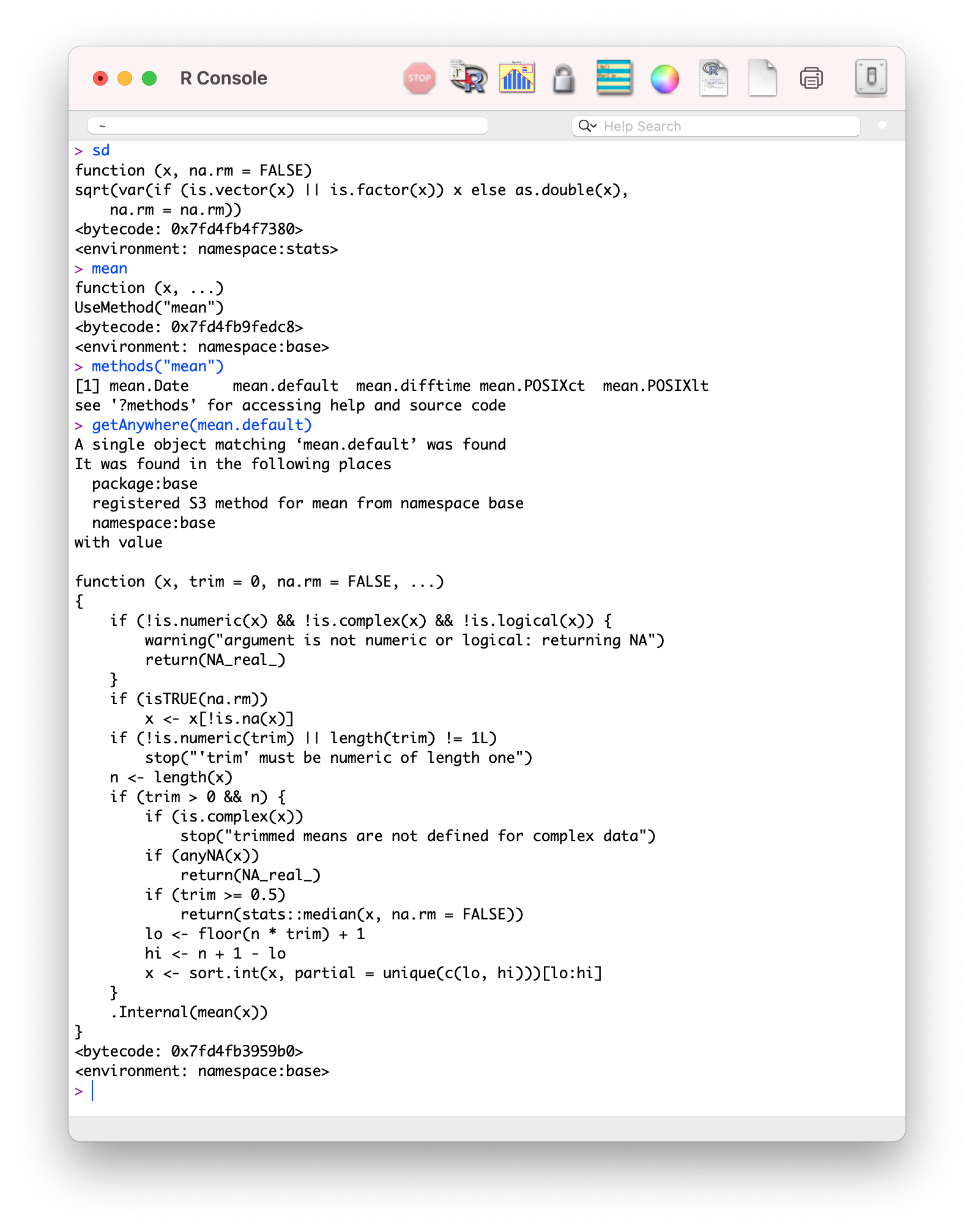}
%\end{adjustwidth}
\caption{How to obtain the code of the functions \texttt{sd}, and \texttt{mean}, which compute the standard deviation and the mean for a set of values, respectively, from library {\it{base}}, and examples of use of \texttt{methods} and \texttt{getAnywhere} .\label{figd}}
\end{figure}

Another advantage to point out, is the fact that we can access the code of the \texttt{R} functions, because it allows us to learn through its reading and reuse them to adapt to new needs or situations. The easiest way to do it is typing the name of the function without parenthesis. For example, in  Figure \ref{figd}, we can see the code of function \texttt{sd} without the comments, which were suppressed to save memory. If we want to access the original source code, we can download it directly from the Comprehensive R Archive Network (CRAN). CRAN is a network of ftp and web servers around the world that store identical, up-to-date, versions of code and documentation for \texttt{R}. \footnote{The folder is \$R\_HOME/src/library/PackageName/R/.}

Sometimes, the function calls a partially hidden one, in a namespace. For example, it happens for the \texttt{mean} function, as we can see in  Figure \ref{figd}. \texttt{UseMethod()} hides the code, but it lets us know whether it is a generic \texttt{S3} function that calls a specific method, appropriate for the class of the object. We can ask for it typing \texttt{methods()} and access it through  \texttt{getAnywhere} function. 

%%%%%%%%%%%%%%%%%%%%%%%%%%%%%%%%%%%%%%%%%%
\subsection{Contributed Packages}
Packages are collections of \texttt{R} functions, data, and compiled code in a well-defined format. \texttt{R} comes with a standard set of packages but it is highly extensible through the use of packages for specific functions and applications. 

Currently, the CRAN packages repository features near 20000 available packages \cite{ref-url0}, and we can obtain the names of all them with  \texttt{rownames(available.packages())}. They can be easily installed typing:

\noindent \texttt{> install.packages(``package\_name'')}

\noindent and pressing Enter to proceded with the installation form a mirror. To start using some installed package, we will need to load 

\noindent \texttt{> library(package\_name)}

We can write our own packages and contribute to CRAN.

All the packages are tested regularly on machines running Debian GNU/Linux, Fedora, macOS (formerly OS X) and Windows.

CRAN Task Views aim to provide some guidance about which packages are relevant for tasks related to certain topics. Currently, there are 42 views available, (Table \ref{tab1}) which can be installed automatically, with the function \texttt{install.views} from package \texttt{ctv}

\noindent \texttt{> install.packages(``ctv'') \# If necessary}

\noindent \texttt{> ctv::install.views(`TeachingStatistics'')}

\begin{table}[H] 
\caption{CRAN task views.\label{tab1}}
%\newcolumntype{C}{>{\centering\arraybackslash}X}
\begin{tabular}{ccc}
\toprule
Agriculture& Bayesian& CausalInference \\ ChemPhys&ClinicalTrials&Cluster \\
Databases& DifferentialEquations & Distributions \\ Econometrics & Environmetrics & Epidemiology\\
ExperimentalDesign & ExtremeValue & Finance \\ FunctionalData & GraphicalModels & HighPerformanceComputing\\
Hydrology & MachineLearning & MedicalImaging \\ MetaAnalysis & MissingData & MixedModels \\
ModelDeployment & NaturalLanguageProcessing & NumericalMathematics \\ OfficialStatistics & Optimization & Pharmacokinetics \\
Phylogenetics & Psychometrics & ReproducibleResearch \\ Robust & Spatial & SpatioTemporal\\
SportsAnalytics & Survival & TeachingStatistics \\ TimeSeries & Tracking & WebTechnologies\\
\bottomrule
\end{tabular}
\end{table}

%%%%%%%%%%%%%%%%%%%%%%%%%%%%%%%%%%%%%%%%%%
\section{Graphical User Interfaces}
To make it easier to use, there have been various attempts to create a more graphical interface, from code editors that interact with \texttt{R} to full GUI's that present the user menus and dialogs, which get things done right away. While they sacrifice the full power of \texttt{R} for ease of usage, the gain in usability may take up for it. The goal of these projects is two fold: encourage non-technical users to learn and perform analyses without programming getting in their way and increase the efficiency of expert R users when performing common tasks by replacing hundreds of keystrokes with a few mouse clicks. Consequently, many data scientists prefer to work in the command line, but beginners should choose a GUI.

\begin{figure}[H]
%\begin{adjustwidth}{-\extralength}{0cm}
\centering
\includegraphics[width=10.5cm]{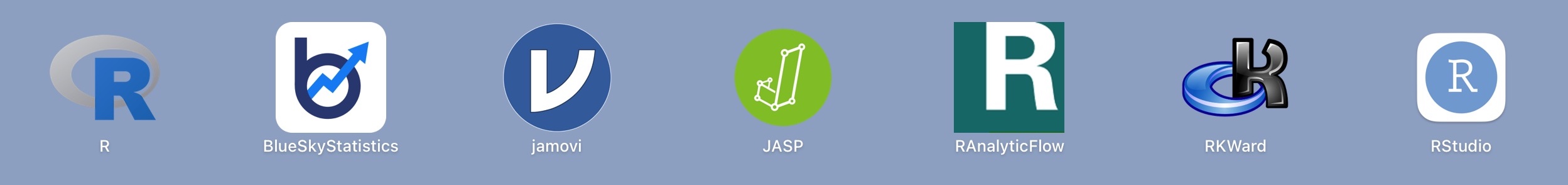}
%\end{adjustwidth}
\caption{Some R GUIs.\label{figb}}
\end{figure}  

We've listed a few GUIs below, with a brief description of each one, so the reader does not waste time searching them.

\begin{description}
\item [BlueSky]  \cite{ref-url4}: free and open sourced fully featured statistics and machine learning application that incorporates comprehensive Six Sigma and Design of Experiments capabilities for Windows and Mac, all accessible through an intuitive and easy-to-use application, without programming required. It shares many of SPSS' features. When we use BlueSky the R code is hidden until we click the button ``</>''. However, for R coding, it uses tidyverse style while most of the other GUIs use base R functions. Another problem we have found when using it is that in the latest version there are two alternatives for some procedures, whose outputs do not match, at least one of them not being correct.
\item [Deducer]  \cite{ref-url5}: designed to be used with the Java based R console JGR, though it supports other R environments (e.g. Windows RGUI and RTerm), as a free and easy to use alternative to proprietary data analysis software such as SPSS, JMP, and Minitab. It has a very nice-looking interface, with a menu system to do common data manipulation and analysis tasks, and an excel-like spreadsheet in which to view and edit data frames. Probably, Deducer is the first R GUI to offer in true APA-style word processing tables, a feature that has been copied by several other GUIs.
\item [Jamovi]  \cite{ref-url6}: free and open statistical software for the desktop and cloud which bridges the gap between researcher and statistician. Jamovi has an extremely interactive interface that shows the result of every selection in each dialog box (as JASP) and saves the setting. We can see the code used at each step, though it uses its own functions from the jmv package. Jamovi's ease of use makes it ideal for introducing people to statistics, and its advanced features ensure students will be well equipped for the rigours of real research when they graduate. Currently, based on our experience, it seems to us to be the best GUI choice for a first course in statistics, and for this reason we will develop it in more detail below.
\item [JASP]  \cite{ref-url7}: intuitive interface which offers standard analysis procedures and more advanced statistical techniques in both their classical and Bayesian approaches. Facilitating Bayesian analysis and a Machine Learning module is its differentiating feature from other GUIs. Unfortunately, JASP does not show the R code below. It is available for Windows, Mac OS X and Linux and we can also download a pre-installed Windows version that will run directly from a USB or external hard drive without the need to install it locally or run JASP in our browser to launch it online via rollApp.
\item [R AnalyticFlow]  \cite{ref-url8}: in addition to intuitive user interface, it also provides advanced features for R experts. They enable to share the processes of data analysis between users with differing levels of proficiency. What sets it apart from other GUIs for R is that it uses a flowchart-like workflow diagram to control analysis instead of just menus. With workflow tools, we get the benefit of the diagram describing the big picture, while the dialog settings in each node control what happens at each step.
\item [R Analytic Tool To Learn Easily (Rattle)]  \cite{ref-url9}: collection of utilities functions for the data scientist, that focuses on beginners looking to point-and-click their way through machine learning tasks. A Gnome (RGtk2) based graphical interface is included with the aim to provide a simple and intuitive introduction to R for data science, allowing the user to quickly load data from a CSV file (or via ODBC), transform and explore the data, build and evaluate models, and export models as PMML (predictive modeling markup language) or as scores. A key aspect of the GUI is that all R commands are logged and commented through the log tab. If the work involves machine learning and/or artificial intelligence instead of standard statistical methods, Rattle is the most adequate GUI \cite{ref-j1}.
\item [R Commander (Rcmdr)]  \cite{ref-url10}: a Basic-Statistics GUI, implemented as an R package, Rcmdr. R Commander allows the users to focus on statistical methods rather than on remembering and formulating R commands. Moreover, by rendering the generated commands visible to users, it has the potential for easing the transition to writing R commands, at least for some interested users. The R code it writes is classic and rarely using the newer tidyverse functions. For this reason, we think it is a very good option for a first course in statistics, especially if students will have a second course in the subject. Its principal problem is that it accesses only a small fraction of the capabilities of R, despite been itself extensible through plug-in packages, many of which are now available on CRAN. 
\item [R-Instant]  \cite{ref-url11}: written in Visual Basic, it is currently available only  for Microsoft Windows. However, a Linux version is under development. R-Instant offers one of the most extensive collections of data wrangling, graphics, and statistical analysis methods of any R GUI. At a basic level, its graphics dialogs are easy to use and at advanced level it offers powerful multi-layer support (as the ggplot2 package). To use its full modeling capabilities, we need to know exactly which {\texttt R} packages we have to use and how their functions work (recognizing a package::function combination is much easier than recalling it without assistance). Therefore, it is not a good choice for beginners.
\item [R KWard]  \cite{ref-url12}: blends a nice point-and-click interface with the IDE most advanced of all the other GUIs reviewed in this work. It is easy to install and start, and it saves all the dialog box settings, allowing the user to rerun them. However, that is done step-by-step, not all at once as Jamovi’s templates allow. The code RKWard creates is classic {\texttt R}, with no tidyverse at all.
\item [RStudio]  \cite{ref-url3}: most trusted IDE for {\texttt R} and Python, available in open source and commercial editions, which runs on the desktop (Windows, Mac, and Linux). We will describe it more deeply in the next section.
\item[Tinn-R] \cite{ref-url14}: generic ASCII/UNICODE text editor/processor for the Windows operating system, well integrated into the R, with GUI and IDE features. None of its features make it stand out from the rest.
\end{description}

Despite \texttt{R}  is a statistical computing environment for data analysis widely adopted by research and industry professionals in STEM, social sciences and humanities and more and more educators use \texttt{R} to teach data analysis, according to our experience, many barriers remain as:
\begin{itemize}
\item The initial difficulty of setting up R may be the first obstacle for novice student.
\item As \texttt{R} is syntax-based, it may be hard to learn for some students, especially those who come in with no prior programming experience.
\item Same students ``hate programming'' and initial feelings of fear toward \texttt{R} might turn them off statistics.
\item For many instructors, just the idea of teaching \texttt{R} may be challenging, particularly when they themselves may have little or no experience with \texttt{R} 

\end{itemize} 
In our opinion, there is no perfect user interface for everyone; each GUI for R has features that appeal to a different set of people, but clearly a good choice of the GUI can break all these barriers. Based on our experience, teaching statistics courses and subjects for more than 20 years to students of different levels and professionals from different areas, we recommend RStudio for statistical computing and Jamovi and/or R-Commander for teaching and applied statistics.

\subsection{Jamovi}
\begin{figure}[H]
%\begin{adjustwidth}{-\extralength}{0cm}
\centering
\includegraphics[width=12.5cm]{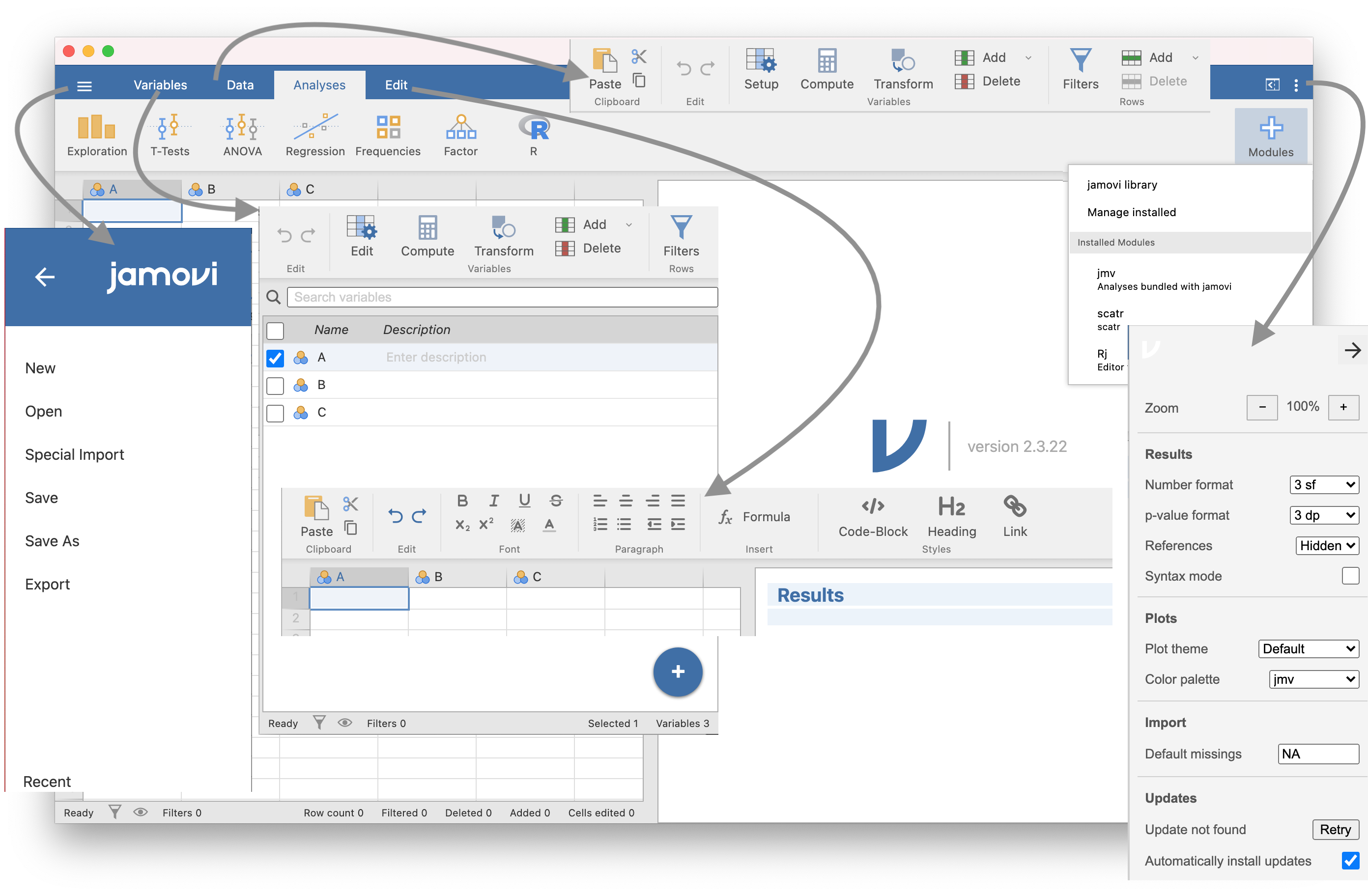}
%\end{adjustwidth}
\caption{Jamovi GUI, with opened windows from tab menu.\label{figJamovi}}
\end{figure}

Jamovi is an advanced real-time statistical spreadsheet, providing a suite of common statistical methods such as descriptives, t-test, ANOVAs, linear and logistical regression, proportion tests, contingency tables and factor analysis. Once installed (it is not necessary to install \texttt{R} previously, like another GUIs, which avoids the first difficulty we mentioned above), when we start Jamovi we will face a user interface which looks something like Figure \ref{figJamovi}. On the left side, there is the spreadsheet view (with each column representing a variable), and the results of statistical analysis appear on the right side.
To perform an analysis, we only have to choose the statistical tool from the Analysis menu and select variables and values that appear in the emerging windows. At the same time, the results will appear on the right side, and they will be updated immediately as we make changes in the options. 

Another interesting feature is that Jamovi also provides an {\it{R Syntax Mode}} which produces equivalent \texttt{R} code for each analysis. All the analyses included with Jamovi are available from within \texttt{R} using the package \texttt{jmv}.

\subsection{R-Commander}
R Commander \cite{ref-url10} provides a GUI for \texttt{R}, based on the \texttt{tcltk} package. Another point-and-click interface loads data and calls \texttt{R} functions to perform the usual analyses involved in introductory Statistics courses. When we start R Commander, the window shown in Figure \ref{figRcmdr} will be opened:
\begin{figure}[H]
%\begin{adjustwidth}{-\extralength}{0cm}
\centering
\includegraphics[width=12.5cm]{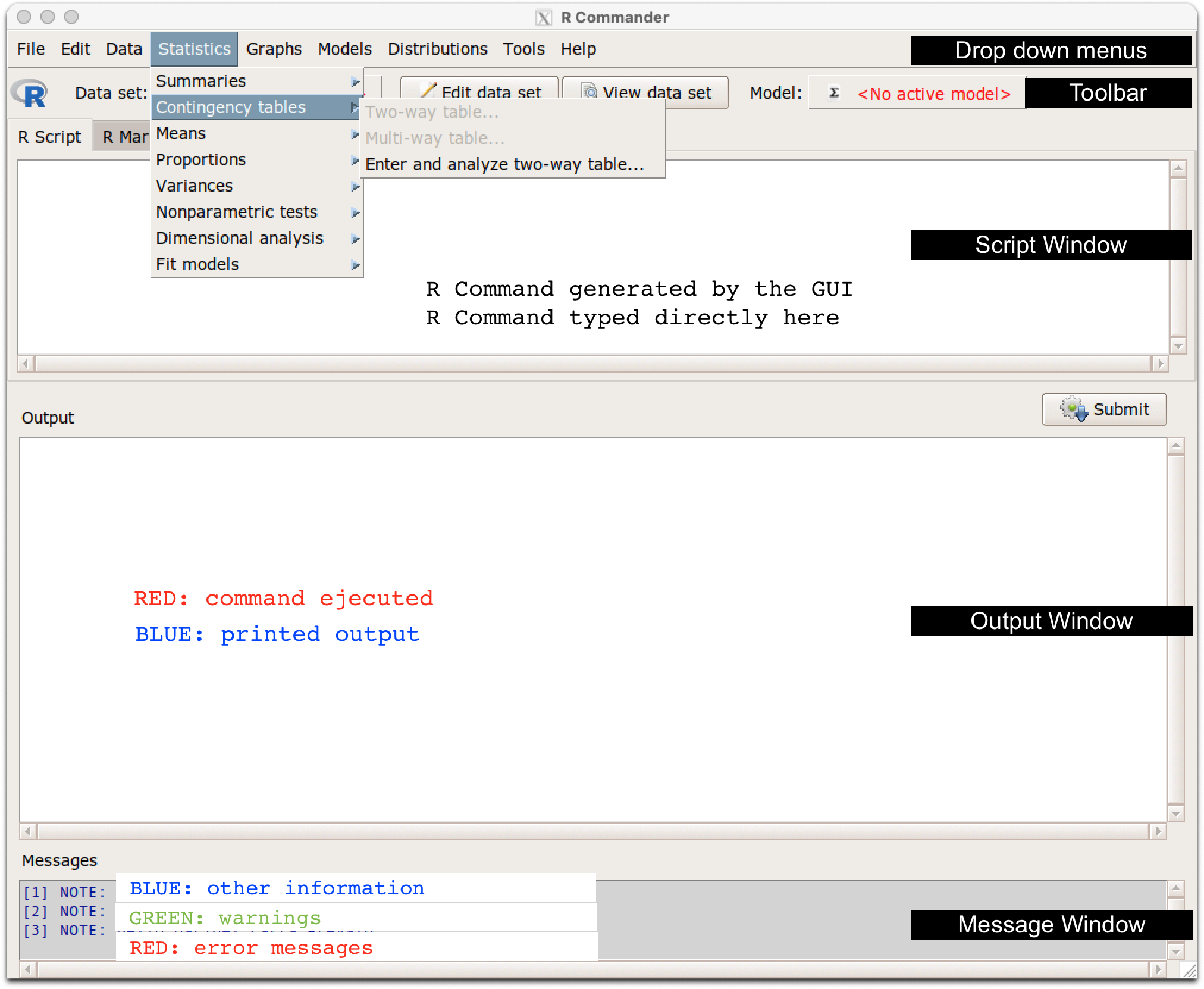}
%\end{adjustwidth}
\caption{Default layout Rcmdr, which is divided into two panes: the source for scripts and Markdown documents (top) and the R Console (bottom)\label{figRcmdr}}
\end{figure}  

Like Jamovi, it has drop down menus that can drive the statistical analysis of data (Figure \ref{figRcmdr}), making it especially useful to \texttt{R} novices. However, R Commander displays the underlying \texttt{R} code for each analysis the user runs, attached to results in the console. Even more, the code appears in the upper script, which allows us to reproduce the analysis, get used to the code of \texttt{R} and start out in programming: which is a feature Jamovi did not facilitate. But, in this case, the changes we make on data or analysis do not update automatically the results, and we have to execute again the corresponding instructions (through menus or compiling script instructions after being modified). R Commander does not generate a report, but a sequence of \texttt{R} code and the outputs.

Most of usual statistical analysis can be made, although more advanced and specialized tools are also available, some of them via plugins, which extend the range of application. We can see their names with:

\noindent \texttt{all.package.names <- rownames(available.packages())}\\
\noindent \texttt{all.package.names[grep("RcmdrPlugin", all.package.names)]}

% \begin{figure}[H]
%\begin{adjustwidth}{-\extralength}{0cm}
%\centering
%\includegraphics[width=15.5cm]{Figures/RcmdrP.png}
%\end{adjustwidth}
%\caption{Rcmdr's Plugins .\label{RcmdrP}}
%\end{figure}  

For more information on the R Commander GUI, see the introductory manual distributed with the package (accessible via the menu: Help $\rightarrow$ Introduction to the R Commander) or the Reference \cite{ref-bookRcmdr}.

\subsection{RStudio}
\begin{figure}[H]
%\begin{adjustwidth}{-\extralength}{0cm}
\centering
\includegraphics[width=6.5cm]{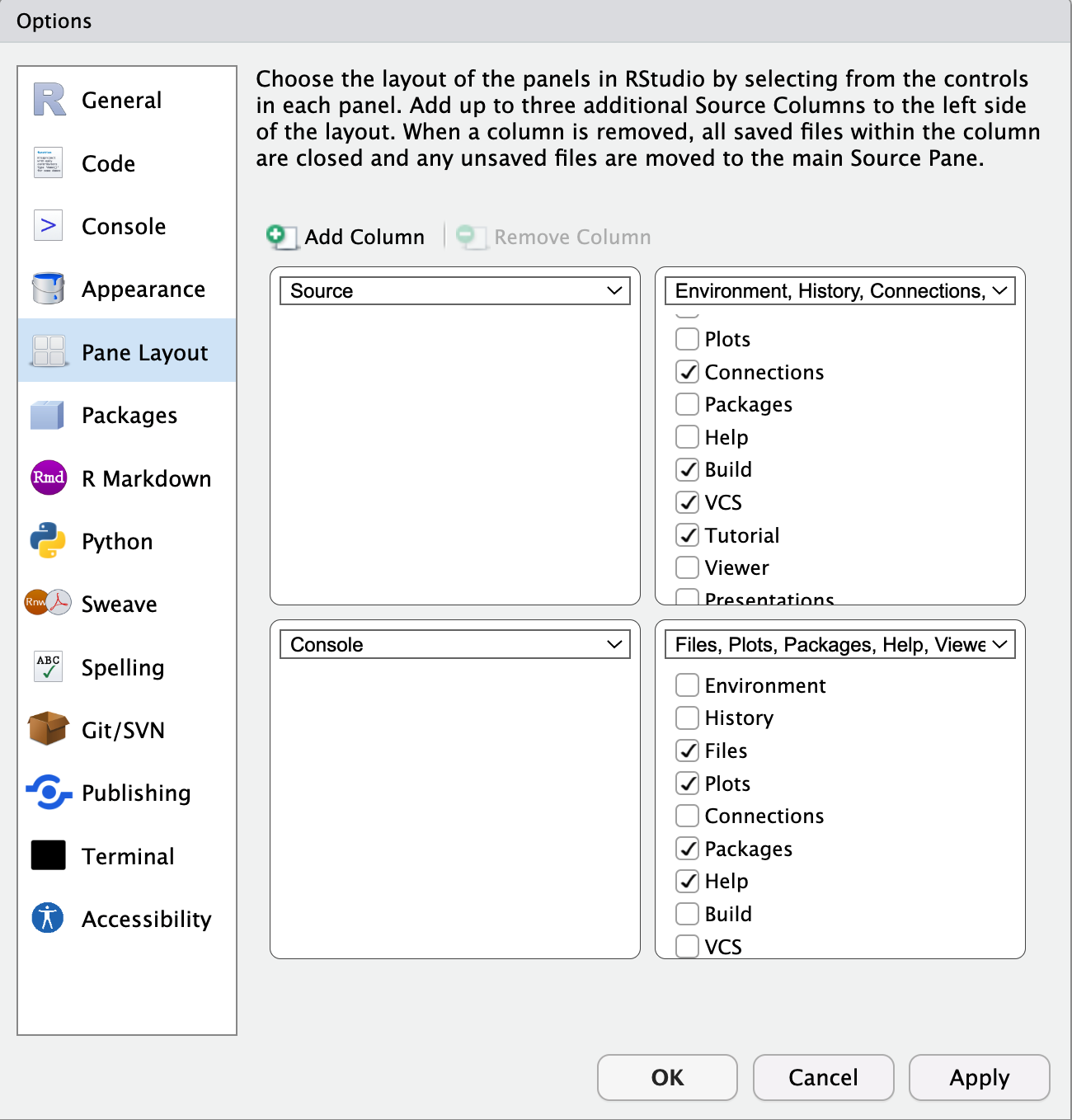}
%\end{adjustwidth}
\caption{How to change the default 4-pane layout.\label{figa1}}
\end{figure}

RStudio \cite{ref-url3} is our favorite example of a code editor that interfaces with R, because it's way ahead of its fellow competitors. The free GUI supports two formats: desktop application, which is the most used one, and server, which runs on a distant server and we can access it with our browser.
% Example of a figure that spans the whole page width. The same concept works for tables, too.
\begin{figure}[H]
%\begin{adjustwidth}{-\extralength}{0cm}
\centering
\includegraphics[width=12.5cm]{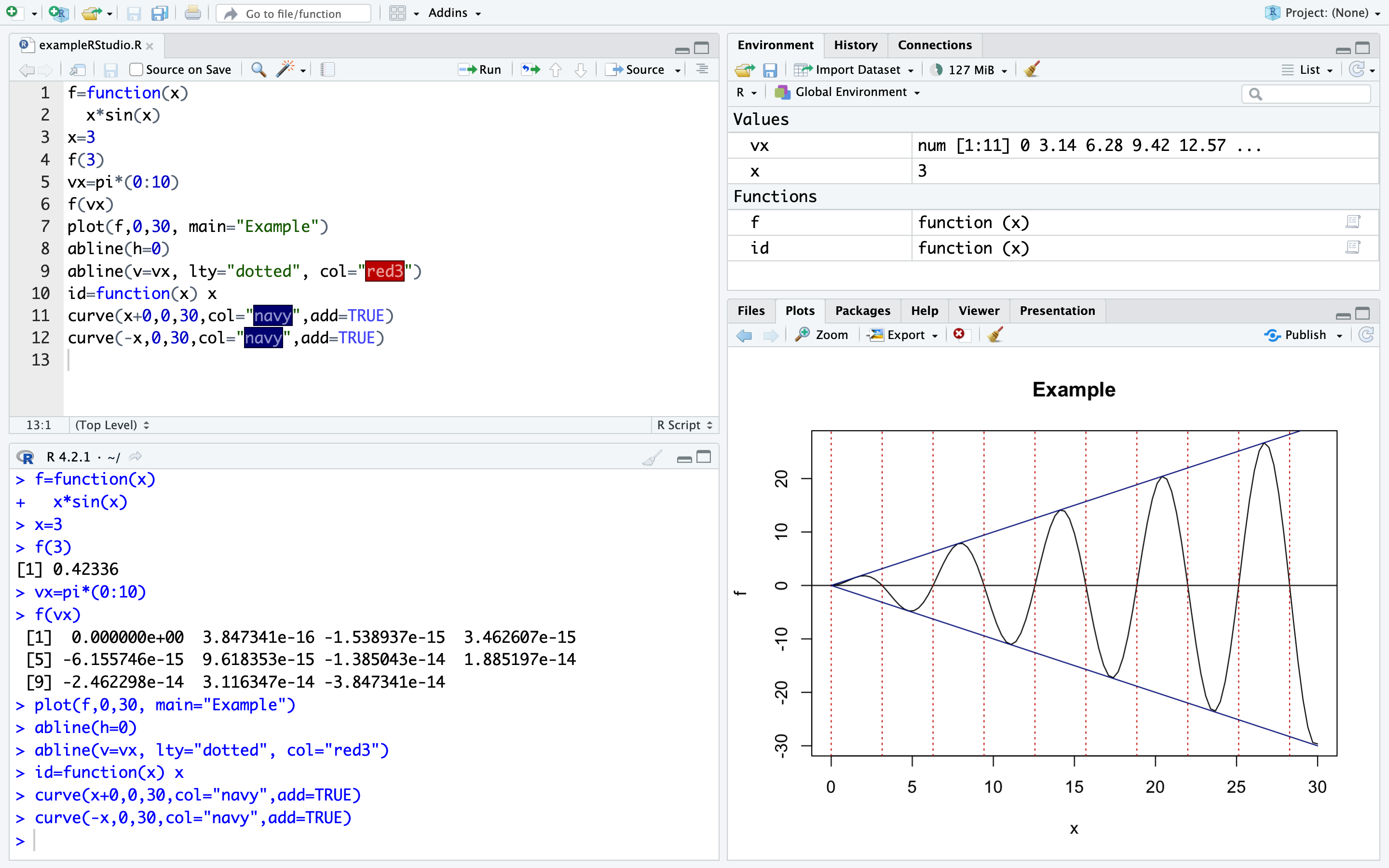}
%\end{adjustwidth}
\caption{Default layout RStudio. It is divided into four panes: the Source for scripts and documents (top-left) that supports direct code execution, the \texttt{R} Console (bottom-left), Environment/History (top-right), and Files/Plots/Packages/Help/Viewer (bottom-right).\label{figa}}
\end{figure}  
With RStudio all the information we need is available in a single window (Figure~\ref{figa}) to write code, navigate the files on our computer, inspect the objets we are created, visualize the generated plots, develop packages, write Shiny apps, ... The placement of these panes and their content can be customized (see menu, Tools $\rightarrow$ Global Options $\rightarrow$ Pane Layout, Figure \ref{figa1}). Additionally, with many shortcuts, code completion, smart indentation editor, syntax highlighting for the major file types you use while developing in \texttt{R}, RStudio will make typing easier and less error-prone. 

\section{Another interesting elements}
The easiness for the use of \texttt{R} go further, both for teaching and research. We will point out and describe briefly some of them:

\subsection{Shiny}
It's an R package that easily enables building interactive web applications that can execute R code on the backend. With shiny, we can host standalone powerful interactive applications on a webpage, embed interactive charts in R Markdown documents, or build dashboards, entirely in R. It can perform any R calculation we can run on our desktop.

Shiny combines the computational power of R with the interactivity of the modern web.

\subsection{RPubs}
\begin{figure}[H]
%\begin{adjustwidth}{-\extralength}{0cm}
\centering
\includegraphics[width=10.5cm]{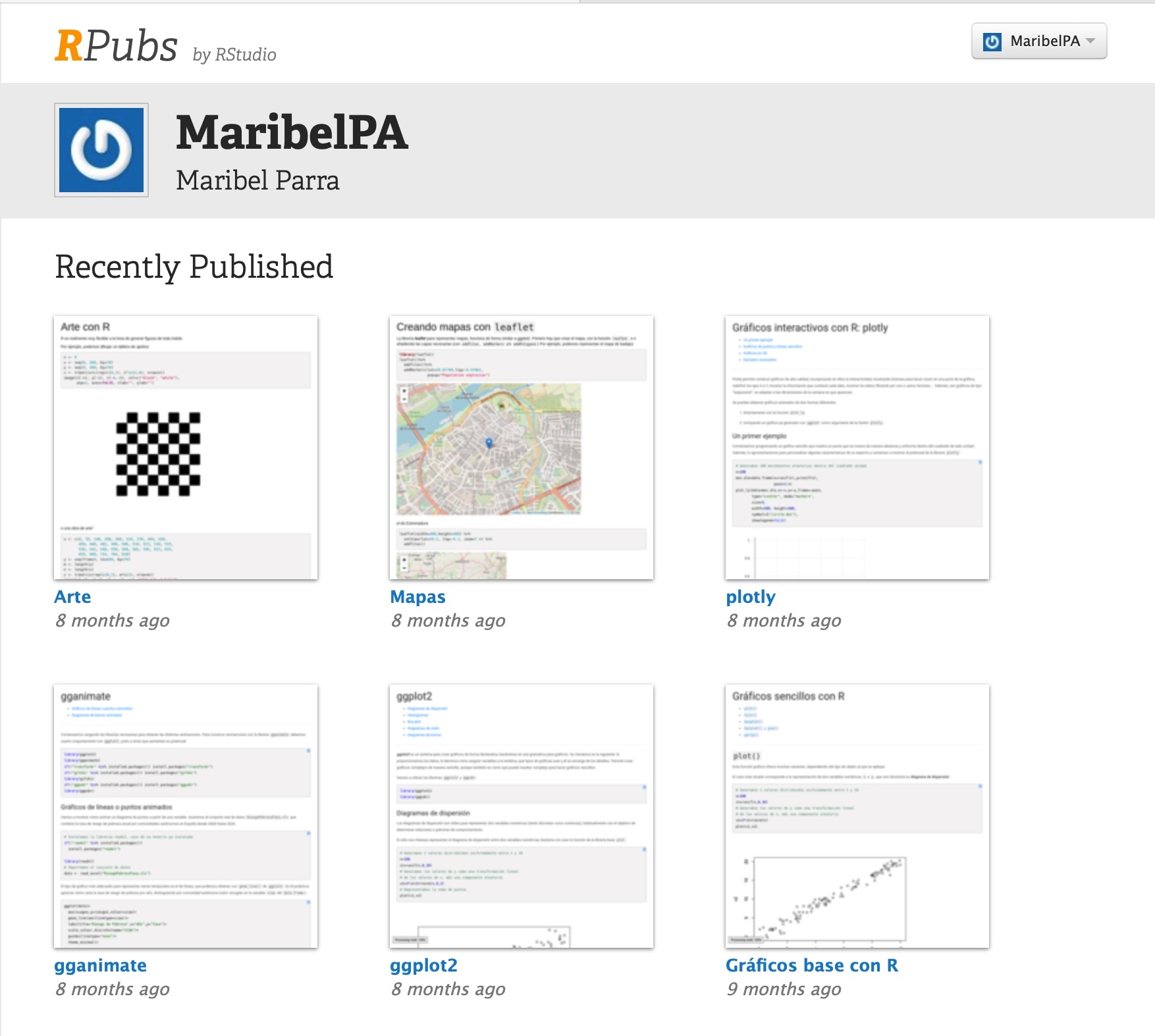}
%\end{adjustwidth}
\caption{Example of profile in RPubs \cite{ref-urlRP}.\label{figRP}}
\end{figure}

RStudio allows us to harness the power of R Markdown to create documents that weave together our writing and the output of R.If we open an account\footnote{Completing \fcolorbox{black}{lightgray}{\text{Register}}} on RPubs \cite{ref-url15}, we can publish those documents on the web with only three steps:
\begin{enumerate}
\item Create a new R Markdown document in RStudio: File$\rightarrow$New$\rightarrow$R Markdown.
\item Click the \fcolorbox{black}{lightgray}{\text{Knit HTML}} button in the doc toolbar, to preview your document.
\item In RStudio, click the \fcolorbox{black}{lightgray}{\text{Publish}} button, in the preview window.
\end{enumerate}
These steps are very intuitive and easy to follow, so it might be the fastest and most comfortable way to write practices and exercises for \texttt{R}
. The success of RPubs has been quite great and it is already possible to find hundreds of interesting practices, worksheets and examples in all the languages (Figure \ref{figRP}).

\subsection{exams}
The package {\it{exams}} provides a one for-all approach to help us to automatically generate, scan and assess the exams, online text and live voting, which are based on exercises (each one is written in a file, either \texttt{.Rmd} or \texttt{.Rnw}) in Markdown or \LaTeX format, possibly including \texttt{R} code, for dynamic generation of exercise elements. These exercises can be of different types: single and multiple choice questions, arithmetic problems, string questions and cloze (combinations thereof). Output formats include standalone files (Docx, HTML, ODT, PDF\footnote{In addition to fully customizable PDF exams, a standardized PDF format (NOPS) is provided that can be printed, scanned, and automatically evaluated.}, ...), Moodle XML, QTI, Blackboard, Canvas, OpenOlat, ILIAS, TestVision, Particify, ARSnova, Kahoot!, Grasple y TCExam.

The site \href{https:/www.r-exams.org/}{\textcolor{cyan}{\underline{R/exams}}} contains different question templates with illustrative examples which are available in different formats of files like markdown, tex or sweave. Exercise templates along with their output and code can be downloaded and inspected as inspiration for new exercises. Each question includes:
\begin{enumerate}
\item Data generating process in \texttt{R}: Programming a good-data-generating process is often the hardest part of authoring a new dynamic exercise, but it is crucial for drawing a large number of random variations
\item Question text
\item Solution text
\item Meta-information: type of question, correct solution, name/label, etc.
\end{enumerate}
The idea is to build a pool of dynamic exercises. The same exercise can be rendered easily into varied output formats. The possible questions type are:
\begin{description}
\item [\fcolorbox{black}{lightgray}{\text{\texttt{schoice}}}] (single choice): select the only correct item out of a list of alternatives, with arbitrary number of shuffled distractors, which can be random numbers and/or typical mistakes.
\item [\fcolorbox{black}{lightgray}{\text{\texttt{mchoice}}}] (multiple choice): select all correct items out of a list of alternatives, with arbitrary number of shuffled true or false statements.
\item [\fcolorbox{black}{lightgray}{\text{\texttt{num}}}] (numeric): compute a single numeric value, within a tolerance interval solving a problems which can be based on some random numbers.
\item [\fcolorbox{black}{lightgray}{\text{\texttt{string}}}] (character string): enter the exactly answer as a character string.
\item [\fcolorbox{black}{lightgray}{\text{\texttt{cloze}}}] (combinations of the above): solve a set of questions combining any of the above types.
\end{description}

R/exams offers different mechanisms for drawing random variations of exams:
\begin{itemize}
\item Random selection of exercises for each student.
\item Random shuffling alternative answers in single and multiple choice questions.
\item Random selection of numbers, text, blocks, graphics, ... using the \texttt{R} programming language, for a range of question formats.
\end{itemize}

\subsection{tidyverse}
Data science is an exciting discipline that allows us to torn raw data into understanding, insight and knowledge. The {\it{tidyverse}} \cite{tidy} is an opinionated collection of \texttt{R} packages designed specifically for data science. They share an underlying design, philosophy, grammar, and data structures. We can install the complete {\it{tidyverse}} with a simple command

\noindent \texttt{> install.packages(``tidyverse'')}

\noindent and run

\noindent \texttt{> library(``tidyverse'')}

\noindent to load the following {\it{tidyverse}}  packages, and make them available in our \texttt{R} session. These packages are called the core, because they are used in almost every analysis:

\begin{description}
\item [\fcolorbox{black}{lightgray}{\text{\texttt{ggplot2}}}]  is a system for declaratively creating graphics, which are based on The Grammar of Graphics.
\item [\fcolorbox{black}{lightgray}{\text{\texttt{dplyr}}}] provides a grammar of data manipulation, solving the most common data manipulation challenges.
\item [\fcolorbox{black}{lightgray}{\text{\texttt{tidyr}}}] includes a set of functions to help us to get tidy data; this is, data with a consistent form (each variable goes in a column, and each column is a variable).
\item [\fcolorbox{black}{lightgray}{\text{\texttt{readr}}}] provides a friendly and fast way to read rectangular data (csv, fwf and tsv).
\item [\fcolorbox{black}{lightgray}{\text{\texttt{purrr}}}]  enhances \texttt{R}'s functional programming toolkit by providing a consistent and complete set of tools for applying functions to element(s) from vector(s) and/or list.
\item [\fcolorbox{black}{lightgray}{\text{\texttt{tibble}}}] is a modern reimagining of the data.frame. Tibbles are lazy and surly; they do less and complain more, forcing to confront problems easier.
\item [\fcolorbox{black}{lightgray}{\text{\texttt{stringr}}}] facilitates the work with strings.
\item [\fcolorbox{black}{lightgray}{\text{\texttt{forcats}}}] provides a set of useful tools that solve common problems with categorical variables or factors. 
\end{description}

\subsection{learnr}
This package allows to turn any Markdown document into an interactive tutorial. These tutorials consist of content along with interactive components (narrative, figures, equations, quiz questions, videos and interactive Shiny components) for checking and reinforcing understanding, and automatically preserve work done within them. If a user works on a few exercises or questions and returns to the tutorial later, he can pick up right where he left off.

\begin{figure}[htp]
%\begin{adjustwidth}{-\extralength}{0cm}
\centering
\includegraphics[width=\textwidth]{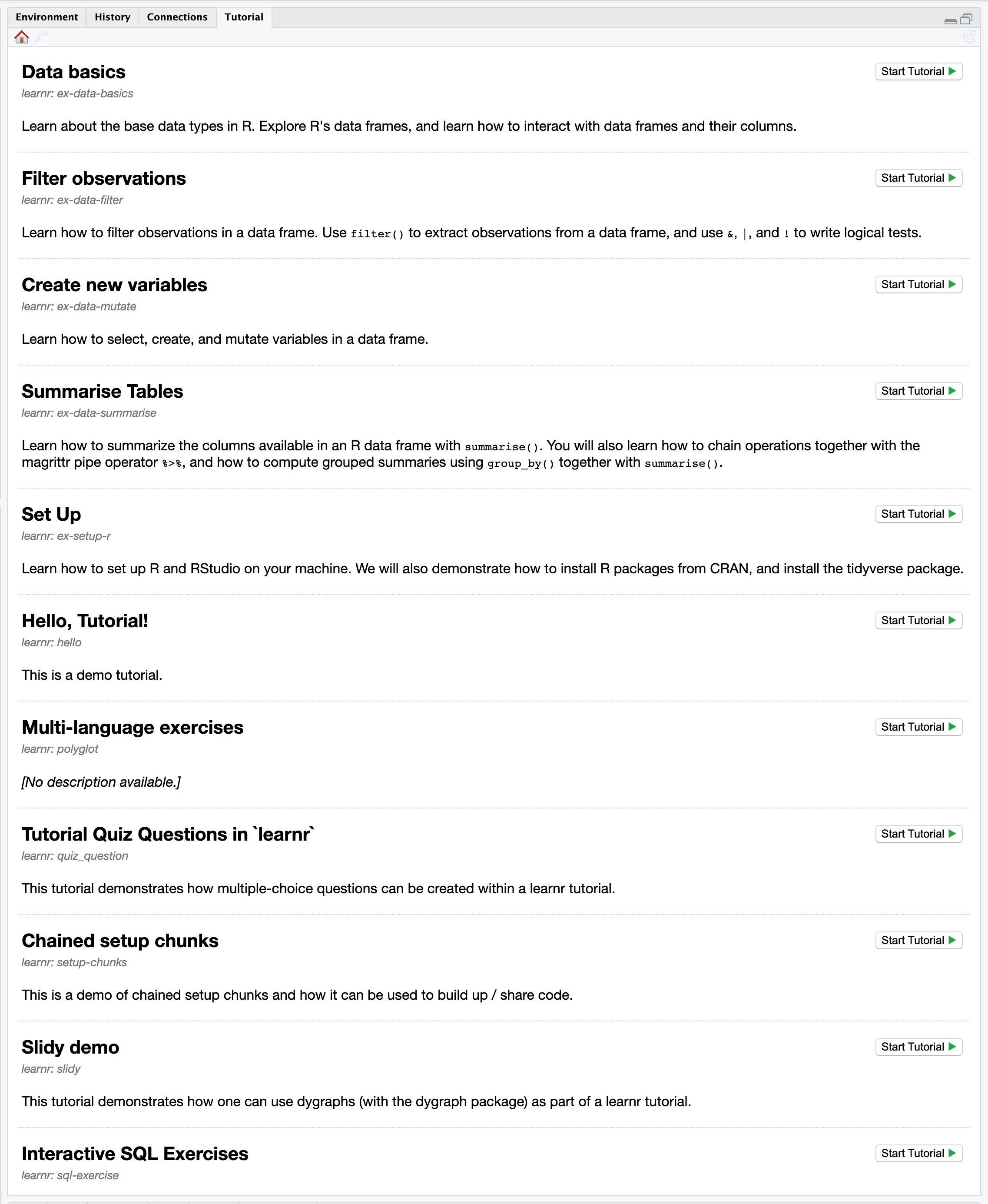}
%\end{adjustwidth}
\caption{RStudio Tutorials powered by the learnr package.\label{figlearnr}}
\end{figure}

Once the package learnr is installed, we can enable the fold Tutorial in any of the right windows on RStudio, and when we click on it we can access the tutorials from learnr (Figure \ref{figlearnr}).

\section{Discussion}

As we mentioned in the Introduction, the employment of Statistics as a tool to analyze data from research has become essential, especially in the last years. For most scientific or technological disciplines, even in many social and humanistic fields, there exists the need of handling basic statistical concepts, in order to be capable of extracting conclusions from the researchers' own data or to understand published research. That is the reason why subjects of Statistics, at different levels, are included in very different academic curriculum, with students owning previous knowledges and capacities in a very diverse range. Besides, it is obvious that each statistical technique requires adequate software to perform calculations and create appealing graph reports.

For all these reasons, there is a growing need for a statistical software as simple and versatile as possible. After more than 20 years of experience as students, teachers and researchers, our choice is \texttt{R}, without any doubts. The more we use it, the bigger advantages we find. We highlight the most important ones:

\begin{itemize}
\item It is available for all the operating systems (Windows, Macintosh and Unix systems).
\item It can interact with another programming languages.
\item It is resource-friendly (it consumes very few system resources).
\item It allows to read data from other softwares such as SPSS, SAS, Excel, etc.
\item It is a very powerful tool for any kind of processing and manipulation of data. 
\item It provides great quality charts, exportable to different formats (bmp, eps, jpeg, pdf, pictex, png, ps, svg, tiff).
\item It includes some advanced and robust techniques that can only be performed with this software.
\item It is regularly updated, and its development has no limits.
\item It provides much information about its functions and packages, which are open-coded.
\item It is very easy to find free resources to learn and use it.
\item It allows to create interactive web applications (apps) with the tool Shiny.
\item It enables to make a work flow to write dynamic and reproducible reports, in several formats (pdf, docx, html).
\end{itemize}

However, the use of a command console instead of a GUI is a big handicap for a beginner. Due to this, the attempts to introduce the use of \texttt{R} in a first course of Statistics, or in Applied Statistics, can lead to bad results, especially for students with no previous programming experience. As an example, we show the learning rates for the students in the course {\it Use of the package exams of R to generate exams or quizzes in Moodle}, framed into the Training Plan for teachers and researchers in the University of Extremadura 2022 (see Table \ref{tab5}).

\begin{table}[H] 
\caption{Learning rates of the course {\it Use of the package exams of R to generate exams or quizzes in Moodle}.\label{tab5}}
\centering
%\newcolumntype{C}{>{\centering\arraybackslash}X}
\begin{tabular}{lccc}
\toprule
\textbf{Field}	& \textbf{Failed} &\textbf{Passed}	&  \textbf{Passed (\%)}\\
\midrule
Arts \& Humanities		& 2		& 0 & 0\%\\
Science 	& 1			& 4 & 80\%\\
Health Sciences		& 2			& 3 & 60\%\\
Social and Legal Sciences	& 3		& 5 & 62.5\% \\
Enginnering and Architecture & 0 & 0 & -\\
\bottomrule
\end{tabular}
\end{table}

\begin{table}[H] 
\caption{Learning rates of the course {\it Data visualization and animation with R}.\label{tab6}}
%\newcolumntype{C}{>{\centering\arraybackslash}X}
\centering
\begin{tabular}{lccc}
\toprule
\textbf{Field}	& \textbf{Failed} &\textbf{Passed}	&  \textbf{Passed (\%)}\\
\midrule
Arts \& Humanities		& 0 & 1 &	100\%		 \\
Science 	& 2 & 9 & 	82\% \\
Health Sciences		& 2			& 2 & 50\%\\
Social and Legal Sciences	& 0 & 9	& 100\%	\\
Enginnering and Architecture & 1 & 2 & 67\%\\
\bottomrule
\end{tabular}
\end{table}
As we can see in the Table \ref{tab5}, the results were not good, especially for teachers not belonging to Science field. 

In contrast, the results obtained for another course of the same Training Plan entitled {\it Data visualization and animation with R}, where we employed RStudio, were better (see Table \ref{tab6}). When we use a GUI, even if it involves some programming, like RStudio, the process of learning can be much more fruitful.   

Therefore, the use of GUIs is ideal for people who only analyze data occasionally, freeing them from the work of writing codes, because they only require to know which statistical tool they need to use, and locate it in menus and dialogue boxes. In this work, we have described the most employed GUIs, showing that one important flaw of most of them (like R-Commander, Deducer, Rattle or Bluesky) is the lack of a key element: reproducibility. Jamovi does not have this problem, and it was designed to look familiar for SPSS users. 

A recent experience of the use of Jamovi in a course took place in November 2022, entitled {\it The code of Science: keys to make research in Health Sciences}, with 16 students. The process of teaching was far more grateful, and the students rated the methodology employed in the course with a mean mark of 4.53 (out of 5), even though they were not familiar with Statistics or Jamovi.

%%%%%%%%%%%%%%%%%%%%%%%%%%%%%%%%%%%%%%%%%%
\section{Conclusions}

\begin{itemize}

\item Despite of having been designed for data analysis, \texttt{R} belongs to the top ranks of most employed programming languages, even for general purposes.

\item \texttt{R} provides all the usual statistical tools in its base package, but also the most specific and modern techniques, together with amazing graphical tools, are available in its nearly 20000 packages.

\item For users with poor or no programming knowledge, GUIs are very helpful. Many of them have been developed, with different purposes. We can highlight the easiness and reproducibility of Jamovi, and the great help R Commander offers to make data analysis in a very simple way, while the user learns programming. 

\item RStudio IDE offers a perfect framework to make the most of \texttt{R} for users with programming skills.

\item Some \texttt{R} packages have become tools of great utility in education. For example, {\it{exams}} allows to create exams and lists of personalized exercises in a very simple way, interacting with Moodle and another educational platforms.
\end{itemize}

%%%%%%%%%%%%%%%%%%%%%%%%%%%%%%%%%%%%%%%%%%
\vspace{6pt} 

%%%%%%%%%%%%%%%%%%%%%%%%%%%%%%%%%%%%%%%%%%
%% optional
%\supplementary{The following supporting information can be downloaded at:  \linksupplementary{s1}, Figure S1: title; Table S1: title; Video S1: title.}


%%%%%%%%%%%%%%%%%%%%%%%%%%%%%%%%%%%%%%%%%%
\subsection*{\textbf{Acknowledgments}}
This research was funded by  MCIN/AEI/10.13039/501100011033 and ERDF A way of making Europe as part of R\&D\&I Project PID2021-122209OB-C32.


\begin{thebibliography}{999}

\bibitem[R(2022)]{ref-url1}
The R Project for Statistical Computing. Available online: \url{https://www.r-project.org} (accessed on 25 May 2023).

\bibitem[Ihaka(1996)]{inicio}
Ihaka, R; Gentleman, R. R: A language for Data Analysis and Graphics. {\em Journal of Computational and Graphical Statistics} {\bf 1996}, {\em 5}(3), 299--314.

\bibitem[R(2022)]{ref-url2}
The R Graph Gallery. Available online: \url{https://r-graph-gallery.com} (accessed on 25 May 2023).

\bibitem[PR(2022)]{ref-url0}
Available Packages. Available online: \url{https://cran.r-project.org/web/packages/} (accessed on 25 May 2023).


\bibitem[RStudio(2022)]{ref-url3}
RStudio is now Posit, our mission continues. Available online: \url{https://posit.co} (accessed on 25 May 2023).

\bibitem[BlueSky(2022)]{ref-url4}
BlueSky Statistics. Available online: \url{https://www.blueskystatistics.com/Default.asp} (accessed on 25 May 2023).

\bibitem[Deducer(2022)]{ref-url5}
Deducer: A GUI for R. Available online: \url{https://www.deducer.org/} (accessed on 25 May 2023).

\bibitem[Jamovi(2022)]{ref-url6}
Jamovi. Available online: \url{https://www.jamovi.org} (accessed on 25 May 2023).

\bibitem[JASP(2022)]{ref-url7}
JASP: A Fresh Way to Do Statistics. Available online: \url{https://jasp-stats.org} (accessed on 25 May 2023).

\bibitem[RAP(2022)]{ref-url8}
R AnalyticFlow: Designed for data analysis. Great for everyone. Available online: \url{https://r.analyticflow.com/en/} (accessed on 25 May 2023).

\bibitem[Rattle(2022)]{ref-url9}
Rattle:  A Graphical User Interface for Data Science in R. Available online: \url{https://rattle.togaware.com} (accessed on 25 May 2023).

\bibitem[GW(2009)]{ref-j1}
Williams, G. Rattle: A Data Mining GUI for R. {\em The R Journal} {\bf 2009}, {\em1}(2), 45-55.

\bibitem[Rcmdr(2022)]{ref-url10}
Rcmdr: R Commander. Available online: \url{https://socialsciences.mcmaster.ca/jfox/Misc/Rcmdr/} (accessed on 25 May 2023).

\bibitem[RInstat(2022)]{ref-url11}
R-Instat. Available online: \url{https://r-instat.org/index.html} (accessed on 25 May 2023).

\bibitem[RKWard(2022)]{ref-url12}
RKWard is an easy to use and easily extensible IDE/GUI for R. Available online: \url{https://rkward.kde.org} (accessed on 25 May 2023).

\bibitem[TinnR(2022)]{ref-url14}
Tinn-R Editor: A modern editor/word processor, perfect for R statistical language. Available online: \url{https://tinn-r.org/en/} (accessed on 25 May 2023).

\bibitem[Rcmdr(2017)]{ref-bookRcmdr}
Fox, J. \textit{Using the R Commander: A Point-and-Click Interface for R}, 1st ed.; Chapman \& Hall/CRC Press: Boca Raton, FL, 2017.

\bibitem[RPubs(2022)]{ref-url15}
RPubs by RStudio: Easy web publishing from R. Available online: \url{https://rpubs.com} (accessed on 25 May 2023).

\bibitem[RP(2022)]{ref-urlRP}
Profile MaribelPA. Available online: \url{https://rpubs.com/MaribelPA} (accessed on 25 May 2023).

\bibitem[Wickham(2017)]{tidy}
Wickham, H.;Golemund, G. \textit{R for Data Science: Import, Tidy Transform, Visualize, and Model Data}, 1st ed.; O'Reilly Media: Sebastopol, CA, 2017.

% Artículos didáctica
\bibitem[Tucker(2022)]{ref-journal}
Tucker, Mary C.;Shaw,  Stacy T.; Son, Ji Y. and Stigler, James W. Teaching Statistics and Data Analysis with R. {\em  Journal of Statistics and Data Science Education} {\bf 2022}, \url{https://www.tandfonline.com/doi/full/10.1080/26939169.2022.2089410}.



\end{thebibliography}
\end{document}